\documentclass[aps,prx,groupaddress,twocolumn,floatfix]{revtex4-1}
\usepackage{units}
\usepackage{amsmath}
\usepackage{amssymb}
\usepackage{graphicx}
\usepackage{bm}
\usepackage{multirow,color,relsize,ulem,microtype}

\newcommand{\be}{\begin{equation}}
\newcommand{\ee}{\end{equation}}
\newcommand{\e}{\varepsilon}

\begin{document}

\title{Chiral symmetry in non-Hermitian systems: product rule and Clifford algebra}

\author{Jose D. H. Rivero}
\affiliation{\textls[-18]{Department of Physics and Astronomy, College of Staten Island, CUNY, Staten Island, NY 10314, USA}}
\affiliation{The Graduate Center, CUNY, New York, NY 10016, USA}

\author{Li Ge}
\email{li.ge@csi.cuny.edu}
\affiliation{\textls[-18]{Department of Physics and Astronomy, College of Staten Island, CUNY, Staten Island, NY 10314, USA}}
\affiliation{The Graduate Center, CUNY, New York, NY 10016, USA}

\date{\today}

\begin{abstract}
Chiral symmetry provides the symmetry protection for a large class of topological edge states. It exists in non-Hermitian systems as well, and the same anti-commutation relation between the Hamiltonian and a linear chiral operator, i.e., $\{H,\Pi\}=0$, now warrants a symmetric spectrum about the origin of the complex energy plane. Here we show two general approaches to identify and generate chiral symmetry in non-Hermitian systems, with an emphasis on lattices with detuned on-site potentials that can vary in both their real and imaginary parts. One approach utilizes the Clifford algebra satisfied by the Dirac matrices, while the other relies on the simultaneous satisfaction of non-Hermitian particle-hole symmetry and bosonic anti-linear symmetry, extended beyond simple spatial transformations to include, for example, an imaginary gauge transformation.
\end{abstract}

\maketitle

\section{Introduction}

Chiral symmetry is pivotal among other symmetries that provide protection to topological states \cite{Hasan,Qi,Alicea,Beenakker,Sarma_RMP,Sarma_QI,Alicea_PRX}. It was originally conceived in quantum mechanics to describe the conserved handedness of Dirac fermion fields, and it often accounts for the symmetry of energy bands with respect to the Fermi level or the energy of an uncoupled orbital in condensed matter physics. A natural extension of chiral symmetry can be introduced in non-Hermitian physics \cite{NPreview,NPhyreview,RMP}, where the energy spectrum is complex in general. Its definition via the anti-commutation relation of the Hamiltonian and a linear operator now warrants a complex spectrum symmetric about the origin of the complex energy plane \cite{Malzard,zeromodeLaser}. As a consequence, a non-Hermitian zero mode, with its energy right at the origin of the complex plane, can still exist similar to its Hermitian counterpart. Such exotic states and associated topological phases of matter have attracted fast-growing interest in photonics and related fields \cite{Lu,St-Jean,Bahari,Bandres,Zhao,Pan,Parto,Leykam}, which are inherently open systems and demand a systematic study and characterization of non-Hermitian chiral symmetry.

On the one hand, the easiest way to construct a non-Hermitian system with chiral symmetry is to maintain the sublattice symmetry of an underlying Hermitian system (such as in a tight-binding square or honeycomb lattice without on-site detunings) and lift its Hermiticity by introducing asymmetric couplings \cite{zeromodeLaser}. While this approach can be applied to both periodic \cite{Malzard} and finite-size systems, it does not utilize one important benefit provided by the non-Hermitian platforms in optics and photonics \cite{NPreview}, namely, the availability and tunability of gain and loss in optical cavities and waveguides, which are represented by an imaginary detuning between different lattice sites. On the other hand, if we directly apply such an imaginary detuning to a Hermitian system with chiral symmetry, its chiral symmetry will be lifted and we often obtain non-Hermitian particle-hole (NHPH) symmetry instead, which results in a spectrum symmetric about the imaginary axis of the complex energy plane \cite{zeromodeLaser,Malzard,Kohmoto,defectState,NHFlatband_PRJ}.

To overcome these obstacles and facilitate the exploration of topological phases of matter in non-Hermitian systems, we propose in this work two general approaches to identify and generate non-Hermitian chiral symmetry. In the first approach, we extend a product rule where chiral symmetry, denoted by $\Pi$ below, results from the simultaneous satisfaction of NHPH symmetry and bosonic anti-linear symmetry. The former is defined similarly to its Hermitian counterpart, i.e., with the Hamiltonian $H$ anti-commuting with an anti-linear operator $\Xi$; the latter is defined as a commutation relation between the Hamiltonian and an anti-linear operator $\Lambda$, with parity-time (PT) symmetry \cite{Bender1} being a prominent example. 

We do note that non-Hermitian chiral symmetry has been observed by combining NHPH symmetry and PT symmetry \cite{zeromodeLaser,Malzard,defectState,Kawabata}. Our goal here is to show that in more general cases, a non-Hermitian bosonic anti-linear symmetry does not involve just simple spatial transformations, such as parity or rotation in a two-dimensional (2D) lattice. Furthermore, NHPH symmetry can also emerge unintentionally in the absence of established mechanisms. Therefore, the resulting non-Hermitian chiral symmetry can exist in a much broader range of non-Hermitian systems, beyond what has been realized previously. Most notably, we examine several systems where $\Lambda$ includes an imaginary gauge transformation \cite{Hatano} as well. Using this approach, we exemplify flexible control of non-Hermitian zero modes at an exceptional point (EP) \cite{EP1,EP2,EP3,EP4,EP5,EP6,EP_CMT,EP_ring}, achieving spatial localization at the center, one corner, or all corners of a square lattice. 

In the second approach, we discuss how the Clifford algebra satisfied by the Dirac matrices can be utilized to discover and analyze non-Hermitian chiral symmetry, independent of NHPH and bosonic anti-linear symmetries. Through the investigation of several examples, we show that chiral symmetry can be maintained or even generated using complex on-site detunings, including in three dimensions (3D). When applied to periodic systems, such constructions provide a convenient scheme to impose non-Hermitian chiral symmetry to known Hermitian topological models. 

\section{Approach I: product rule}

We first review two important concepts in the fundamental proposition regarding possible forms of symmetries in quantum systems, i.e., the Wigner theorem \cite{Weinberg}. It states that any symmetry transformation is necessarily represented by a linear (and unitary) or anti-linear (and anti-unitary) transformation of the Hilbert space. A linear symmetry transformation ${\cal U}$ satisfies
\be
{\cal U}(a\phi_1+b\phi_2) = a\, {\cal U}\phi_1+b\,{\cal U}\phi_2,
\ee
where $\phi_{1,2}$ are two arbitrary quantum states and the complex numbers $a,b$ are their linear superposition coefficients. In contrast, an anti-linear symmetry operator ${\cal A}$ satisfies
\be
{\cal A}(a\phi_1+b\phi_2) = a^*{\cal A}\phi_1+b^*{\cal A}\phi_2,
\ee
where the asterisks denote the complex conjugation as usual. From this definition, it can be inferred that an anti-linear operator can be represented by the product of a linear operator and the complex conjugation.

%Additionally, according to the standard definition of inner product in Hermitian quantum mechanics, unitary operators satisfy $(\mathcal{U}\psi_1,\mathcal{U}\psi_2) = (\psi_1,\psi_2)$, whereas anti-unitary operators follow $(\mathcal{A}\psi_1,\mathcal{A}\psi_2) \equiv (\psi_2,\psi_1) = (\psi_1,\psi_2)^*$ for any two vectors in the Hilbert space. The implementation of unitary and antiunitary transformations on Hilbert spaces might vary according to the definition of inner product, which is biorthogonal in non-Hermitian systems, leading to the necessity of a generalization of the Wigner theorem, to embrace non-Hermitian systems as well.

As mentioned in the introduction, the first approach we employ to generate non-Hermitian chiral symmetry relies on the simultaneous satisfaction of NHPH symmetry and a non-Hermitian bosonic anti-linear symmetry:
\be
\{H,\Xi\}=0,\quad[H,\Lambda]=0.\label{eq:NHPH_BA}
\ee
Bosonic anti-linear symmetry can be implemented conveniently using strategically placed photonic elements with balanced optical gain and loss \cite{NPreview}.
Meanwhile, a probably more important and intriguing foundation of this approach is that imposing any arbitrary imaginary on-site potentials to an underlying Hermitian chiral lattice with real-valued couplings gives rise to NHPH symmetry automatically \cite{zeromodeLaser}. Therefore, NHPH symmetry can coexist nicely with bosonic anti-linear symmetry enabled by optical gain and loss, which in turn warrants non-Hermitian chiral symmetry as we show in detail below.

Since both $\Xi$ and $\Lambda$ in Eq.~(\ref{eq:NHPH_BA}) are anti-linear operators, they can be written as the product of a linear operator and the complex conjugation $K$:
\be
\Xi \equiv CK,\quad \Lambda \equiv XK.\label{eq:symmetries}
\ee
$K$ is often the manifestation of time-reversal operator for a finite-sized system \cite{Bender1}, and $C$ for the NHPH symmetry mentioned above is given by the chiral operator of the underlying Hermitian lattice, i.e., $C=P_A-P_B$ as in the Su-Schrieffer-Heeger (SSH) model \cite{SSH}, where $P_{A,B}$ are the projection operators onto the two sublattices. These two sublattices are defined such that there is no coupling between two sites on the same sublattice. $X$, on the other hand, can take a variety of forms. For example, two common choices of $X$ in 2D are mirror reflection and rotation, which lead to PT \cite{Bender1,NPreview,NPhyreview} and rotation-time (RT) symmetry \cite{scattering_PRJ,PT_PRX14,conservation2D}, respectively.

As a consequence of Eq.~(\ref{eq:NHPH_BA}), the following symmetry transformations hold for the eigenstates of $H$:
\be
\Xi\psi_\mu=\psi_\nu,\quad\Lambda\psi_\nu=\psi_{\nu'},
\ee
where the subscripts $\mu,\nu,\nu'$ are not necessarily the same. The corresponding energy eigenvalues satisfy
\be
\e_\mu = -\e_\nu^*, \quad\e_\nu = \e_{\nu'}^*,\label{eq:spectrum_doubleSymm}
\ee
i.e., they are symmetric about the imaginary and real energy axis of the complex energy plane, respectively. It is then straightforward to see
\be
\Lambda\Xi\psi_\mu = \psi_{\nu'},\quad \e_\mu = -\e_{\nu'}, \label{eq:prod}
\ee
which indicates the existence of chiral symmetry, i.e., $\{H,\Pi\}=0$, with the \textit{linear} operator $\Pi\equiv\Lambda\Xi=XC^*$. A non-Hermitian zero mode occurs when all the subscripts are the same, leading to $\e_\mu=0$.

This product rule is similar in construction to how chiral symmetry in a Hermitian system can be generated as the product of particle-hole symmetry and time-reversal symmetry \cite{Hasan}. However, we note that since the non-Hermitian chiral operator $\Pi$ is no longer given by the difference of the sublattice projection operators, the wave function of a non-Hermitian zero mode with $\e_\mu=0$ does not necessarily vanish on one sublattice. More importantly, this zero energy can be an EP, i.e., a non-Hermitian degeneracy where two or more eigenstates coalesce \cite{EP1,EP2,EP3,EP4,EP5,EP6,EP_CMT,EP_ring}.

As mentioned in the introduction, this product rule has been implemented in systems where the bosonic anti-linear symmetry involves a simple spatial transformation, such as parity or rotation that is represented by a permutation in the matrix form. However, this product rule is much more general, and in some cases the bosonic anti-linear symmetry, as well as the NHPH symmetry, is complicated and often hidden.

\begin{figure}[t]
\includegraphics[clip,width=\linewidth]{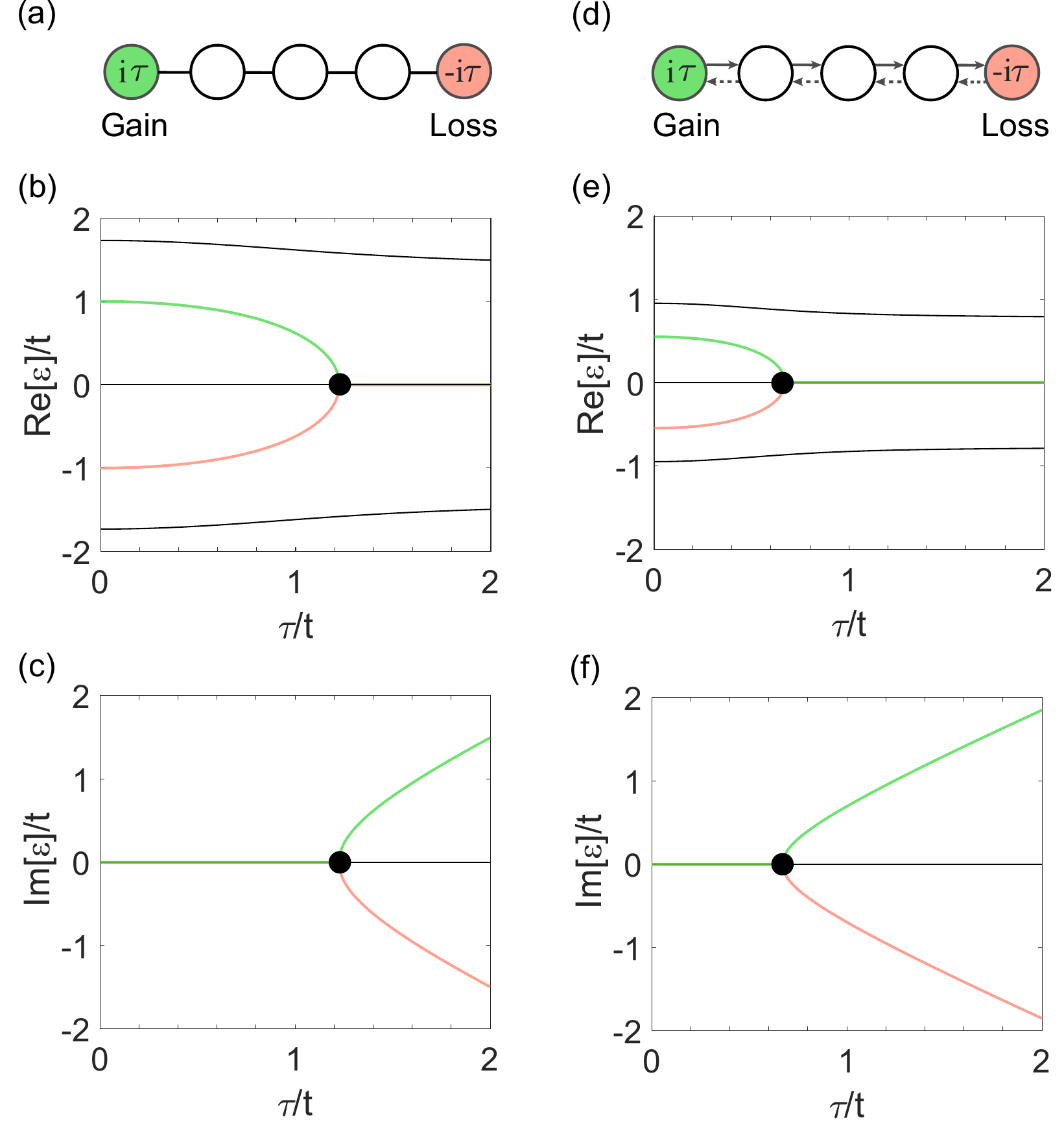}
\caption{\textbf{Generating non-Hermitian chiral symmetry by the product rule}. (a,d) Schematics of two tight-binding lattices. Solid and dashed links indicate couplings $t\in\mathbb{R}$ and $t'=0.3t$. (b,c) Real and imaginary parts of the complex spectrum for the lattice in (a) as a function of the gain and loss strength $\tau$. Black dot marks its exceptional point. (e,f) Same as (b,c) but for the lattice in (d).
} \label{fig:gauge}
\end{figure}

As an example, let us contrast two one-dimensional lattices shown in Figs.~\ref{fig:gauge}(a) and (d). They have the same pair of gain and loss of the same strength (i.e., with imaginary on-site potentials $\pm i\tau$) at the two ends, but system I has symmetric coupling while system II has asymmetric couplings. These lattices can be realized using optical microcavities and their counterparts in microwaves, acoustics and other related fields \cite{NPreview}. When $\tau$ is zero, both systems have sublattice symmetry and hence chiral symmetry as well. When $\tau$ becomes nonzero, system I acquires NHPH symmetry as specified in Eq.~(\ref{eq:symmetries}) and PT symmetry with the parity operator $\cal P$ performing a horizontal mirror reflection. Therefore, it displays a symmetric spectrum  about the origin in the complex energy plane [Figs.~\ref{fig:gauge}(b) and (c)], manifesting its non-Hermitian chiral symmetry explained by the product rule given by Eq.~(\ref{eq:prod}). While the same route to NHPH symmetry still applies in system II with $\Xi=CK$, system II clearly lacks PT symmetry due to the asymmetric couplings. Therefore, it is quite remarkable that the system still displays a symmetric spectrum in the complex energy plane [Figs.~\ref{fig:gauge}(e) and (f)].

To understand this behavior, we note that asymmetric couplings can be regarded as a consequence of an imaginary gauge transformation \cite{Hatano}. However, previous investigations of such gauge transformations have excluded explicitly gain and loss, which by far is the most viable approach to realize non-Hermitian systems in photonics and related areas \cite{NPreview}. Here in the presence of the gain and loss cavity in system II, the imaginary gauge transformation, given by
\be
{\psi_n} = e^{-\frac{1-n}{2}\ln\frac{t}{t'}}\tilde{\psi}_n \equiv s^{n-1}\tilde{\psi}_n, \label{eq:gauge}
\ee
from a system with symmetric coupling $\tilde{t}=\sqrt{tt'}$ and wave function $\tilde{\psi}_n$ in the $n$th cavity, \textit{preserves} the strength of gain and loss. As a result, the operator $X$ in the bosonic anti-linear symmetry is given by
\be
X = G\mathcal{P}G^{-1},\quad G=\mathrm{diag}(1,\,s,\,s^2, \ldots),
\ee
where $\mathcal{P}$ is the same mirror reflection as in system I. Clearly, $X$ is a linear operator and $X^2=G\mathcal{P}^2G^{-1}=\bm{1}$, where $\bm{1}$ is the identity matrix. Together with the aforementioned NHPH symmetry, the non-Hermitian chiral symmetry of system II is then given by $\Pi=XC=G\mathcal{P}G^{-1}C=G\mathcal{P}CG^{-1}$, where we have used $[G^{-1},C]=0$ for these two diagonal operators.

\begin{figure}[t]
\includegraphics[clip,width=\linewidth]{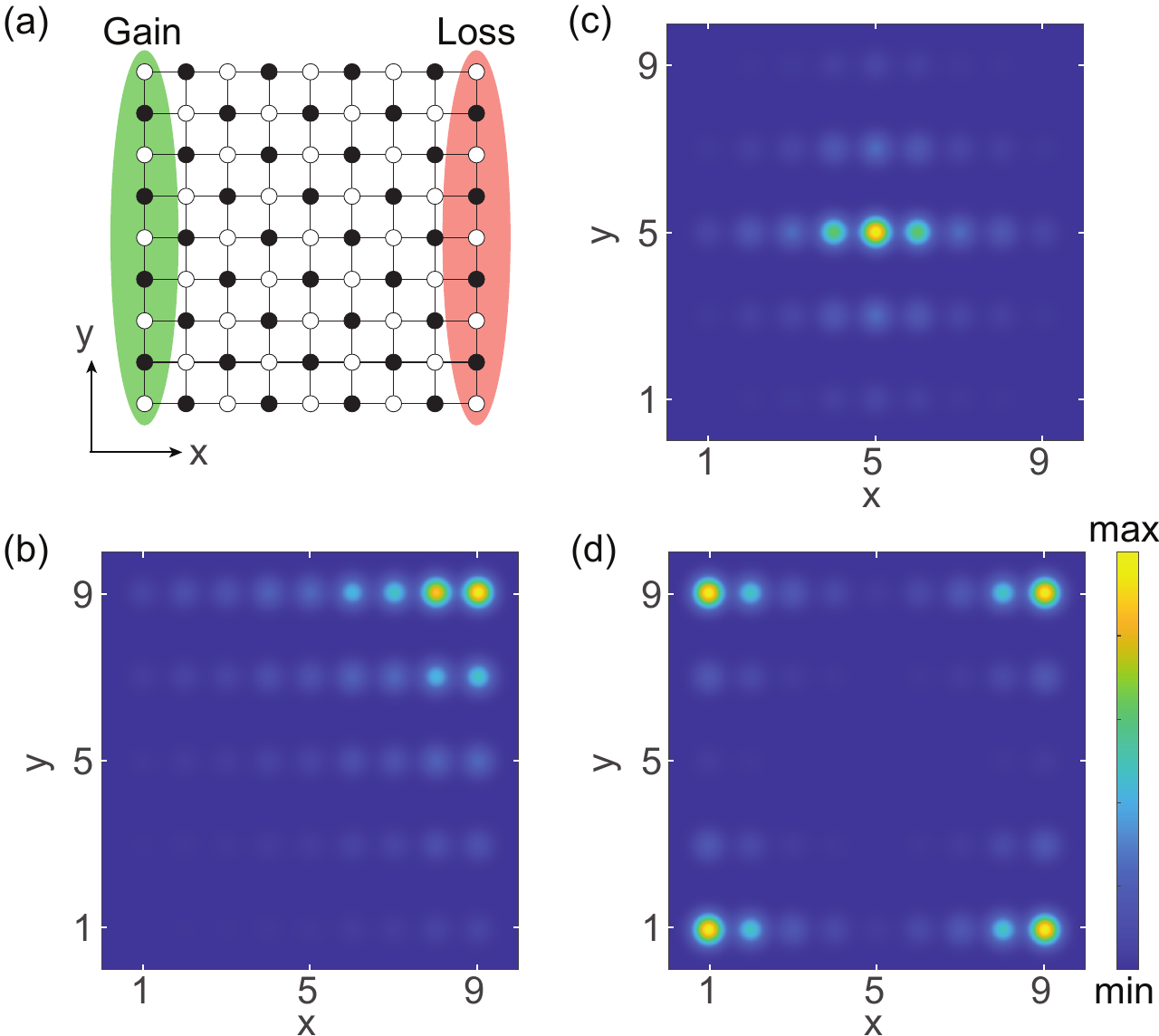}
\caption{\textbf{Localized non-Hermitian zero mode at an EP}. (a) Schematic of a square lattice before an imaginary gauge transformation is applied. Horizontal and vertical couplings are given by $t$ and $1.1t$, respectively. (b--d) Spatial profile of a zero mode at an EP with a modified coupling $t'=0.7t$, $0.5t$ and $0.4t$, respectively. See the main text for the application of this imaginary gauge transformation. 
} \label{fig:gauge2d}
\end{figure}

The imaginary gauge transformation given by Eq.~(\ref{eq:gauge}) is characterized a phase $\phi=i\frac{1-n}{2}\ln\frac{t}{t'}$ that is linear in space. This property leads to asymmetric couplings that are homogeneous in space, i.e., $t$ and $t'$ in system II. More generally, the gauge transformation can involve a phase with a more complicated spatial dependence. In particular, one can localize a zero mode that is at an EP and protected by non-Hermitian chiral symmetry anywhere in the system with ease. Fig.~\ref{fig:gauge2d}(a) shows one example in a square lattice with nine rows and columns, and gain and loss are imposed in the leftmost and rightmost columns. To apply the imaginary gauge transformation, we let the asymmetric horizontal and vertical couplings be $t,t'$ and $1.1t,1.1t'$, respectively. In Fig.~\ref{fig:gauge2d}(b), these asymmetric couplings are still homogeneous in space, and an EP of order 3 \cite{Graefe,Flatband_PT,Loncar,sensingEP3} is reached at $\tau=0.79t$, similar to those in Fig.~\ref{fig:gauge}. The corresponding wave function is exponentially localized at the upper right corner. If instead, we exchange the two asymmetric couplings in the $x$ $(y)$ direction in the right (bottom) half of the lattice, the EP is realized at $\tau=0.94t$ instead, and the non-Hermitian zero mode is localized at the defect of the gauge transformation in both directions, i.e., right at the middle of the square lattice [Fig.~\ref{fig:gauge2d}(c)]. Finally, if we reverse the direction of all these asymmetric couplings, now the non-Hermitian zero mode at its EP is localized at all four corners [Fig.~\ref{fig:gauge2d}(d)], reached when $\tau=0.7t$.

\begin{figure}[t]
\includegraphics[clip,width=\linewidth]{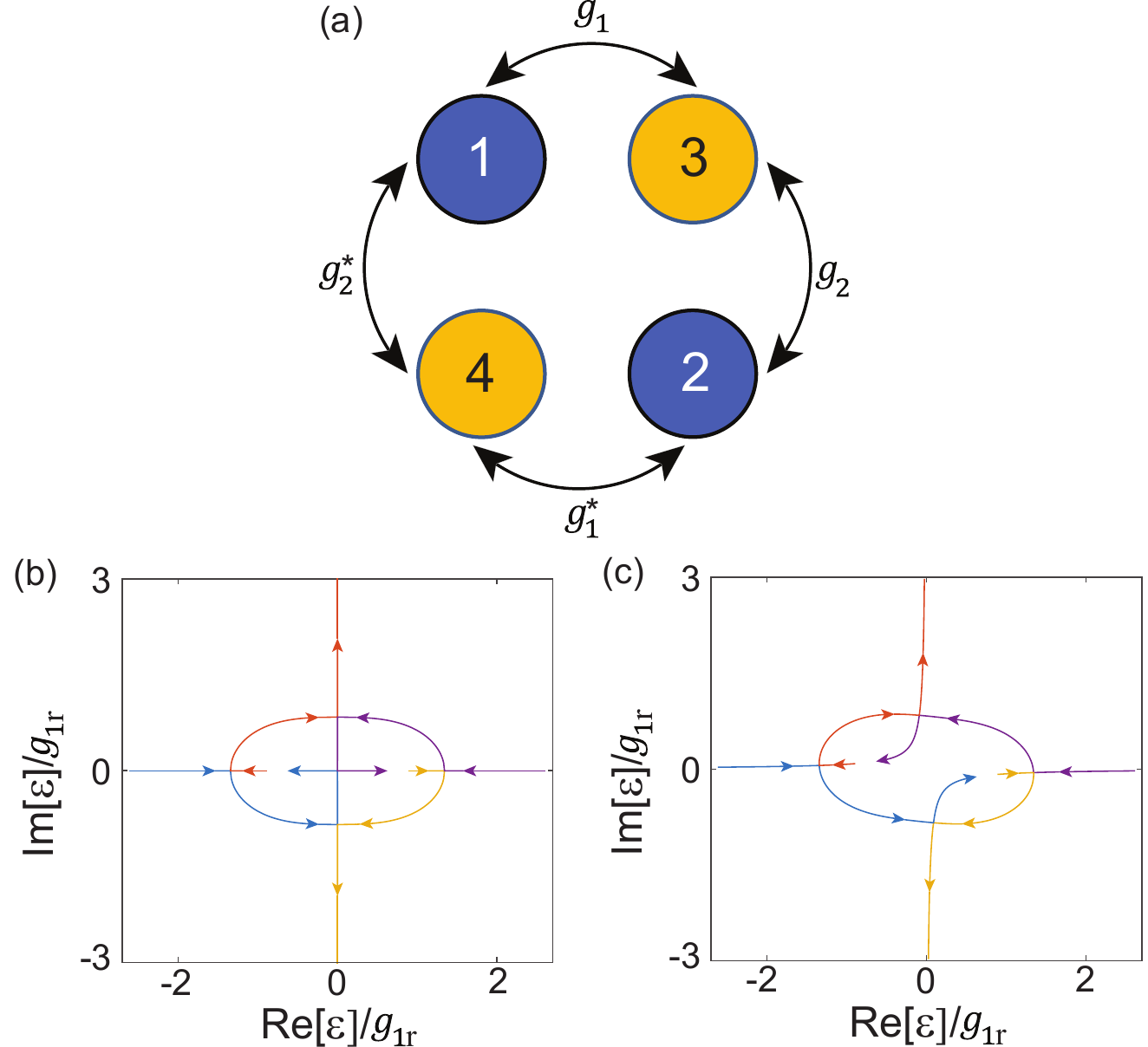}
\caption{\textbf{A hidden non-Hermitian chiral symmetry}. (a) Schematic of a 4-cavity ring with complex couplings $g_1\equiv g_{1r}+ig_{1i}$, $g_2\equiv g_{2r}+ig_{2i}$ and their complex conjugates. Purple and orange denote the on-site detunings $\pm\beta$. (b) Trajectories of the complex energy spectrum when $g_{1i}=g_{2i}$ is increased from 0 to 2$g_{1r}\in\mathbb{R}$. $g_{2r}=2\beta=1.5g_{1r}$ are used. (c) Same as (b) but with $2\beta=(1.5-0.2i)g_{1r}$.} \label{fig:RTwheel}
\end{figure}

In these examples, the imaginary gauge transformation does not affect NHPH symmetry: the transformed lattices by the imaginary gauge still consist of two sublattices with real couplings, and the imaginary on-site detunings due to gain and loss again lead to NHPH symmetry \cite{zeromodeLaser}. Therefore, one just needs to analyze the spatial dependence of the phase $\phi$ to identify $\Xi$ accountable for its bosonic anti-linear symmetry. This task becomes increasingly more difficult with the system size if the variation of $\phi$ is complex or even random. In the meanwhile, there are other non-Hermitian systems where it is NHPH symmetry that is obscure. For example, one may accidentally construct systems with NHPH and non-Hermitian chiral symmetries \cite{Zhou} and be unaware of their symmetry operators. One such case is given in Fig.~\ref{fig:RTwheel}(a), where four sites on a tight-binding ring are coupled by two pairs of complex couplings. This model can be considered as a generalized Rice-Mele model \cite{Rice-Mele}, and its spectrum displays symmetries about the real axis, imaginary axis, and the origin in the complex plane [Fig.~\ref{fig:RTwheel}(b)].

The first property is the result of RT symmetry: the system is invariant under a combined $\pi$-rotation and time-reversal operation when the detunings $\pm\beta$ are real. The NHPH symmetry, on the other hand, is difficult to identify and cannot be revealed using a gauge transformation. To pin down this accidentally generated NHPH symmetry and the resulting non-Hermitian chiral symmetry, below we change the perspective and resort to the second approach mentioned in the introduction, i.e., the Clifford algebra and the Dirac matrices.

\section{Approach 2: Clifford algebra}

The Dirac matrices $\{\gamma^0,\gamma^1,\gamma^2,\gamma^3\}$, also known as the gamma matrices, appear in the Dirac equation to describe relativistic quantum mechanics. They are given by 
\be
\gamma^0 =
\begin{pmatrix}
\bm{1}_2 & 0 \\
0 & -\bm{1}_2
\end{pmatrix}
,
\;
\gamma^j =
\begin{pmatrix}
0 & \sigma_j \\
-\sigma_j & 0
\end{pmatrix}
\;(j=1,2,3)
\ee
in terms of the identity matrix and Pauli matrices. Together with
\be
\gamma^5 =
\begin{pmatrix}
0 & \bm{1}_2 \\
\bm{1}_2 & 0
\end{pmatrix},
\ee
they satisfy the Clifford algebra
\be
\{\gamma^\mu,\gamma^\nu\}=0\;\; (\mu\neq\nu),\quad
\{\gamma^\mu,\gamma^\mu\}=2\xi^{\mu}\bm{1}_4,\label{eq:clifford}
\ee
where $\xi^{\mu}=1$ for $\mu=0,5$ and $-1$ for $\mu=1,2,3$. This defining property of the Clifford algebra is particularly appealing in the generation of non-Hermitian chiral symmetry: by defining the Hamiltonian as a superposition of the Dirac matrices and their products, we have a straightforward way to determine its chiral operators.

For example, if the Hamiltonian includes a linear superposition of individual Dirac matrices, we can enumerate all its chiral symmetries by using Eq.~(\ref{eq:clifford}) as well as
\be
\{g_i\gamma^i + g_j\gamma^j,\xi^ig_j\gamma^i - \xi^jg_i\gamma^j\}=0\quad(i\neq j)\label{eq:clifford1}
\ee
and
\be
\{\gamma^j\gamma^{k},\gamma^{l}\} = 0,\label{eq:anti_3a}
\ee
where $j\neq k$ and $l=j$ or $k$. %(otherwise $\{\gamma^j\gamma^{k},\gamma^{l}\} = 2\gamma^j\gamma^k\gamma^l$)
If the Hamiltonian also contains the product of two Dirac matrices, we may also need to utilize
\be
\{\gamma^j\gamma^k,\gamma^k\gamma^l\}=0\;\;(j\neq l\neq k).\label{eq:anti_4}
\ee
In this analysis, we note that on-site detunings can be expressed by $\gamma^0=\mathrm{diag}(1,1,-1,-1)$, $i\gamma^1\gamma^2=\mathrm{diag}(1,-1,1,$ $-1)$, $\gamma^3\gamma^5=\mathrm{diag}(1,-1,-1,1)$, and the trivial uniform detuning $\bm{1}_4$, as well as their linear superpositions.

Now let us revisit the system shown in Fig.~\ref{fig:RTwheel}(a). Its Hamiltonian can be written as
\be
H = \beta\gamma^0 + g_{1r}\gamma^5 + \gamma^0(g_{2r}\gamma^1+ig_{1i}\gamma^3) + g_{2i}\gamma^2,\label{eq:H_RTwheel}
\ee
where $g_1\equiv g_{1r}+ig_{1i}$ and $g_2\equiv g_{2r}+ig_{2i}$. In the absence of detuning (i.e., $\beta=0$), we find $\gamma^0$ as a chiral operator, which is identical to the sublattice operator $C$ in Eq.~(\ref{eq:symmetries}). Note however, it is not the only chiral operator in this case. Using a generalization of Eq.~(\ref{eq:anti_3a}), i.e.,
\be
\{\gamma^j\tilde\gamma,\tilde\gamma\} = 0,\quad\tilde\gamma=\sum_{k\neq j}a_k\gamma^k,\label{eq:anti_3b}
\ee
we find another chiral operator given by
\be
\Pi = g_{2r}\gamma^1 + ig_{1i}\gamma^3 \label{eq:chiral_RTwheel}
\ee
with proper normalization.

The non-Hermitian chiral symmetry defined by this $\Pi$ operator holds even when $\beta$ is \textit{finite}, whereas that defined by $\gamma^0$ (i.e., the sublattice symmetry) is lifted. $\Pi$ then warrants the spectrum symmetry about the origin of the complex energy plane observed in Fig.~\ref{fig:RTwheel}(b). In the case that $\beta$ is real, the system also has rotation-time symmetry as mentioned. From the perspective of the product rule discussed previously, i.e., $\Pi=\Lambda\Xi$, one can then identify the obscure NHPH symmetry as
\be
\Xi={\cal R}_2K(g_{2r}\gamma^1 + ig_{1i}\gamma^3)={\cal R}_2(g_{2r}\gamma^1 - ig_{1i}\gamma^3)K.\label{eq:NHPH_RTwheel}
\ee
This NHPH symmetry, as well as the RT symmetry, is lifted with a complex $\beta$. However, the non-Hermitian chiral symmetry persists thanks to the Clifford algebra [Fig.~\ref{fig:RTwheel}(c)].

\begin{figure}[t]
\includegraphics[clip,width=\linewidth]{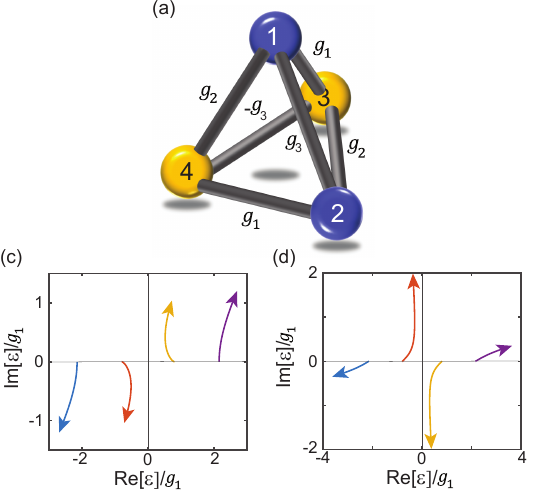}
\caption{\textbf{Non-Hermitian chiral symmetry of a 3D pyramid}. (a) Schematic of the system. Purple and orange denote the on-site detunings $\pm\beta$. (b) Its eigenvalue spectrum with $g_2=g_1\in\mathbb{R}$, $g_3=0.8g_1$ and detuning $\alpha e^{i\pi/4}\gamma^3\gamma^5$, where $\alpha$ is increased from 0 to 2. (c) Same as (d) but with an additional detuning $\alpha e^{i\pi/3}\gamma^1\gamma^2$.}\label{fig:pyramid}
\end{figure}

In the example above, the sublattice symmetry is lifted by a finite detuning. It also disappears with next nearest neighbor (NNN) couplings [i.e., between cavities of the same color arranged diagonally in Fig.~\ref{fig:RTwheel}(a)], with which the system is effectively turned into a pyramid. To show that the Clifford algebra can still lead to non-Hermitian chiral symmetries without two sublattices and their chiral operator $\gamma^0$, we consider a lattice with such connectivity in Fig.~\ref{fig:pyramid}(a) and the following Hamiltonian:
\be
H = g_1\gamma^5 + g_2\gamma^0\gamma^1 + g_3\gamma^1\gamma^5, \label{eq:clifford3}
\ee
where $g_3$ represents the two NNN terms. Its chiral symmetries are specified by $\Pi_1=\gamma^1$ and $\Pi_2=\gamma^0\gamma^5$, which can be checked using Eqs.~(\ref{eq:anti_3a}) and (\ref{eq:anti_4}). $\Pi_1$ tolerates detunings of the forms $\gamma^0$ and $\gamma^1\gamma^2$, shown explicitly below Eq.~(\ref{eq:anti_4}). At the same time,  $\Pi_2$ accommodates detunings given by $\gamma^0$ and $\gamma^3\gamma^5$ [Fig.~\ref{fig:pyramid}(c)]. At first glance, a detuning consisting of a superposition of $\gamma^3\gamma^5$ and $\gamma^1\gamma^2$ (i.e., $H\rightarrow H+d_1\gamma^1\gamma^2 + d_2\gamma^3\gamma^5$) would then lift all chiral symmetries of the system. However, the chiral symmetry given by $\Pi_1$ actually evolves with the detuning, i.e.,
\be
\Pi_1\to\gamma^1+d_2\gamma^1\gamma^3,
\ee
and the non-Hermitian Hamiltonian still has a symmetric spectrum about the origin of the complex energy plane [Fig.~\ref{fig:pyramid}(d)]. Only by having all three forms of detuning can this chiral symmetry be removed.

\begin{figure}[b]
\includegraphics[clip,width=\linewidth]{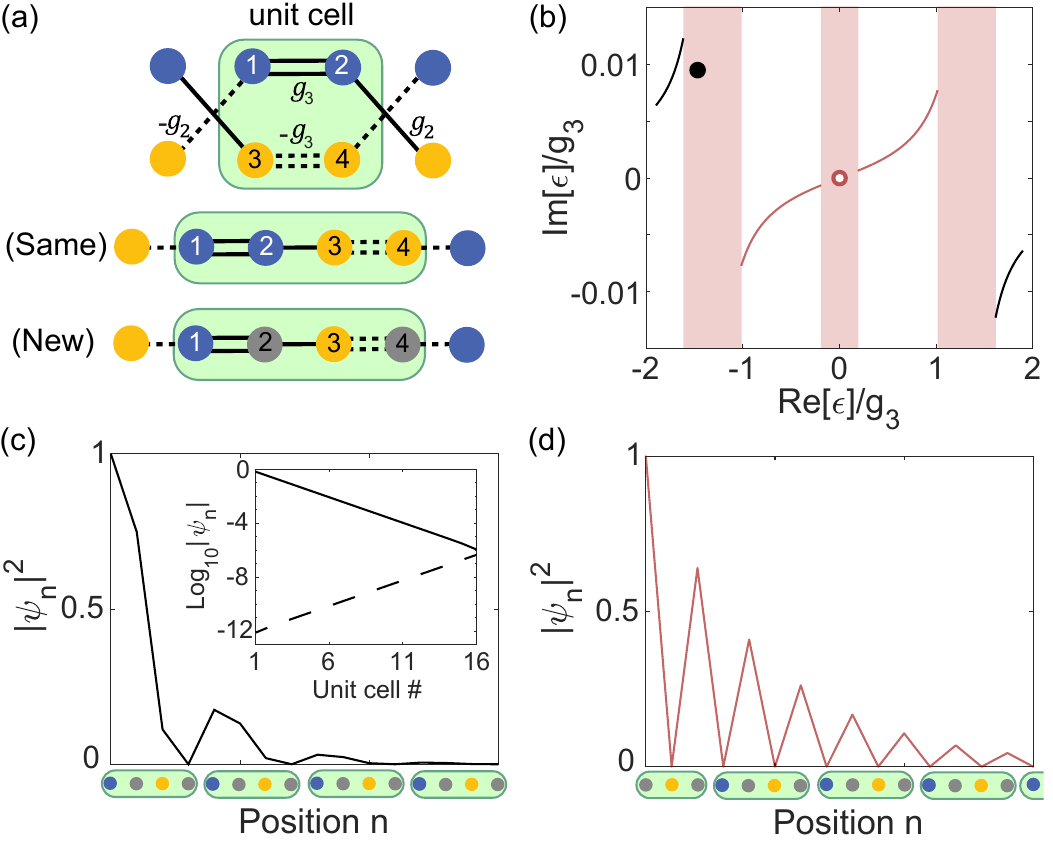}
\caption{\textbf{Non-Hermitian extension of a topological lattice}. (a) Schematic of $H_s$ given by Eq.~(\ref{eq:H_squareRoot}) (top) and its equivalent form (middle). The bottom shows a similar lattice but with different detunings, and its complex band structure (solid lines) is shown in (b). Shaded areas represent the projection of the bulk band gaps on the real energy axis. Closed and open dots in (b) indicate the energy of the edge states in (c) and (d), respectively. They are finite in length with 64 and 63 sites, where purple, orange, and gray denote the on-site detunings $\pm\beta$ and 0. Inset in (c): Amplitude of this edge state in the first (solid) and last (dashed) cavities of all 16 unit cells. $g_2/g_3=0.8$ and $\beta/g_3=-0.3+0.01i$ are used.}\label{fig:squareRoot}
\end{figure}

As an application of this approach to generate non-Hermitian chiral symmetry, we exemplify non-Hermitian extensions of known topological models using the Clifford algebra. The Bloch Hamiltonian of the Hermitian bow-tie lattice considered in Ref.~\citenum{squareRoot} can be written as
\be
H_s = (\beta\gamma^0 + ig_2 \gamma^1\sin k - ig_2\gamma^2\cos k )+ g_3\gamma^1\gamma^5,\label{eq:H_squareRoot}
\ee
where $\beta,g_{1,2}$ are all real couplings and $k$ is the lattice wave vector. $H_s$ actually describes two copies of the same lattice (see Fig.~\ref{fig:squareRoot}), but it is easier to analyze than the Bloch Hamiltonian of the latter using the Clifford algebra. The previously identified chiral symmetry of $H_s$ is given by $\gamma^5$, which can be readily verified using Eqs.~(\ref{eq:clifford}) and (\ref{eq:anti_3a}). Using Eqs.~(\ref{eq:anti_4}) and (\ref{eq:anti_3b}), we also identify another non-Hermitian chiral symmetry
\be
\Pi_2 = \frac{\gamma^1(\beta\gamma^0 - ig_2\gamma^2\cos k)}{N^{\frac{1}{2}}},\quad N = \beta^2+g^2_2\cos^2k
\ee
which is not found in Ref.~\cite{squareRoot}. Here the product of the two chiral operators leads to a linear symmetry, i.e.,
\be
\Pi_2\Pi_1=-\frac{i\beta\gamma^2+g_2\gamma^0\cos k}{N^{\frac{1}{2}}}\equiv W,\quad [H,W]=0,
\ee
and $\{1,\Pi_1,\Pi_2,W\}$ form the Klein group, which is the direct product of two $Z_2$ groups. 

Besides $\gamma^0$, the simple form of $\Pi_1$ also accommodates detunings of the form $\gamma^3\gamma^5$, which is given by $\mathrm{diag}(1,-1,-1,1)$ as mentioned previously. By itself, this form of detunings is equivalent to $\gamma^0$, which is most obvious once we perform a gauge transformation on the wave function in the fourth cavity, i.e., $\psi_4\to-\psi_4$; it flips the sign of all negative couplings, the difference between $\gamma^0$ and $\gamma^3\gamma^5$ lies only in the order of $g_2$ and $g_3$, and the lattice reduces to the SSH model without this form of detunings. 

More importantly, the Clifford algebra analysis leads to the following two observations. First, the chiral symmetry given by $\Pi_1$ persists when these detunings are complex, which offers a straightforward route to study the non-Hermitian extension of this model. Second, these different forms of detunings can coexist without destroying this chiral symmetry. For example, even though the detunings given by $\gamma^0$ and $\gamma^3\gamma^5$ are equivalent as mentioned, their superposition gives a different form of detuning [e.g., $\beta(\gamma^0+\gamma^3\gamma^5)=\mathrm{diag}(2\beta,0,-2\beta,0)$ as depicted in the bottom panel of Fig. 5(a)] and leads to a previously unexplored lattice. By taking $\beta$ to be complex, we observe a complex spectral with non-Hermitian chiral symmetry [Fig.~\ref{fig:squareRoot}(b)]. Due to the ambiguity of defining topological numbers in non-Hermitian systems, here we avoid this discussion and show instead the evidence of topological protection, i.e., edge states in the band gaps of the system. Depending on the terminations at the two ends of the lattice, this lattice can display, for example, an edge state in the left band gap [filled dot in Fig.~\ref{fig:squareRoot}(b)] or right at the origin [open dot in Fig.~\ref{fig:squareRoot}(b)], i.e., a non-Hermitian zero mode. Their spatial profiles are shown in Figs.~\ref{fig:squareRoot}(c) and (d), respectively. While both of them seem to have vanished amplitude in a subset of cavities, only the zero mode is truly dark on one of the original sublattices before the detunings are introduced, consisting of the first and third cavities of each unit cell. The alternate detunings on this sublattice do not affect the spatial profile of this mode, and hence its energy remains at zero. In contrast, the edge state shown in Fig.~\ref{fig:squareRoot}(c) has a ``reflected'' tail: the amplitude of the wave function in the first three cavities of each unit cell attenuates exponentially from left to right, while that in the fourth cavity increases exponentially (see the inset). 
\vspace{4mm}

\section{Conclusion and Discussion}

In summary, we have presented two general approaches to construct systems with non-Hermitian chiral symmetry, aiming to facilitate the exploration of topological phases of matter in non-Hermitian systems, especially on optical and photonic platforms. The first approach relies on the simultaneous satisfaction of NHPH symmetry and non-Hermitian bosonic anti-linear symmetry. We have shown that by going beyond simple spatial transformations such as parity or rotation, a much broader range of non-Hermitian systems can display chiral symmetry, including those with an imaginary gauge transformation. The second approach utilizes the Clifford algebra, and the examples we have discussed are based on the Dirac matrices. They have helped us reveal non-Hermitian chiral and other symmetries whose operators would otherwise remain obscure. Using this approach, we have also investigated the non-Hermitian extension of a known topological model, and we have shown that chiral symmetry as well as topological edge states can persist with complex on-site potentials. Generalizations to more complicated or even higher-dimensional systems can also be achieved, by working with suitable Clifford algebras. 

As a final clarification, we note that chiral symmetry in optics and photonics can also refer to the symmetry between clockwise and counterclockwise modes of motion \cite{Redding,Sarma,Cao,Liu}, which should be distinguished from our discussions here.

This project is supported by the NSF under Grant No. DMR-1506987 and PHY-1847240.


\begin{thebibliography}{99}


%topological quantum computing

\bibitem{Hasan} M. Z. Hasan and C. L. Kane, ``Topological insulators," Rev. Mod. Phys. \textbf{82}, 3045 (2010).
\bibitem{Qi} X.-L. Qi and S.-C. Zhang, ``Topological insulators and superconductors," Rev. Mod. Phys. \textbf{83}, 1057 (2011).
\bibitem{Alicea} J. Alicea, ``New directions in the pursuit of Majorana fermions in solid state systems," Rep. Prog. Phys. \textbf{75}, 076501 (2012).
\bibitem{Beenakker} C. W. J. Beenakker, ``Random-matrix theory of Majorana fermions and topological superconductors," Rev. Mod. Phys. \textbf{87},  1037 (2015).

\bibitem{Sarma_RMP} C. Nayak, S. H. Simon, A. Stern, M. Freedman, and S. Das Sarma, ``Non-Abelian anyons and topological quantum computation," Rev. Mod. Phys. \textbf{80}, 1083 (2008).
\bibitem{Sarma_QI} S. D. Sarma, M. Freedman, and C. Nayak, ``Majorana zero modes and topological quantum computation," npj Quantum Information \textbf{1}, 15001 (2015).
\bibitem{Alicea_PRX} D. Aasen, M. Hell, R. V. Mishmash, A. Higginbotham, J. Danon, M. Leijnse, T. S. Jespersen, J. A. Folk, C. M. Marcus, K. Flensberg, and J. Alicea ``Milestones toward Majorana-based quantum computing," Phys. Rev. X \textbf{6}, 031016 (2016).

\bibitem{NPreview} L. Feng, R. El-Ganainy, and L. Ge, ``Non-Hermitian photonics based on parity-time symmetry,"
Nat. Photon. \textbf{11}, 752--762 (2017).

\bibitem{NPhyreview} R. El-Ganainy, K. G. Makris, M. Khajavikhan, Z. H. Musslimani, S. Rotter, and D. N. Christodoulides, ``Non-Hermitian physics and PT symmetry,"
    Nat. Phys. \textbf{14}, 11--19 (2018).

\bibitem{RMP} V. V. Konotop, J. Yang, and D. A. Zezyulin, ``Nonlinear waves in PT-symmetric systems,"
Rev. Mod. Phys. \textbf{88}, 035002 (2016).

%\bibitem{Moiseyev_book} N. Moiseyev, {\it Non-Hermitian Quantum Mechanics} (Cambridge, New York, 2011).

\bibitem{zeromodeLaser} L. Ge, ``Symmetry-protected zero-mode laser with a tunable spatial profile,"
Phys. Rev. A \textbf{95}, 023812 (2017).
\bibitem{Malzard} S. Malzard, C. Poli, and H. Schomerus, ``Topologically protected defect states in open photonic systems with non-Hermitian charge-conjugation and parity-time symmetry," Phys. Rev. Lett. \textbf{115}, 200402 (2015).

\bibitem{Lu} L. Lu, J. D. Joannopoulos, and M. Soljacic, ``Topological photonics," Nat. Photon. \textbf{8}, 821 (2014).

\bibitem{St-Jean} P. St-Jean et al., ``Lasing in topological edge states of a one-dimensional lattice," Nat. Photon. \textbf{11}, 651--656 (2017).

\bibitem{Bahari} B. Bahari et al., ``Nonreciprocal lasing in topological cavities of arbitrary geometries," Science \textbf{358}, 636--640 (2017).

\bibitem{Bandres} M. A. Bandres et al., ``Topological insulator laser: Experiments," Science \textbf{359}, eaar4005 (2018).

\bibitem{Zhao} H. Zhao et al., ``Topological hybrid silicon microlasers,"
Nat. Commun. \textbf{9}, 981 (2018).

\bibitem{Pan} M. Pan, H. Zhao, P. Miao, S. Longhi, and L. Feng, ``Photonic zero mode in a parity-time symmetric lattice," Nat. Commun. \textbf{9}, 1308 (2018).

\bibitem{Parto} M. Parto et al., ``Complex Edge-State Phase Transitions in 1D Topological Laser Arrays," Phy. Rev. Lett. \textbf{120}, 113901 (2018).

\bibitem{Leykam} D. Leykam, K. Y. Bliokh, C. Huang, Y. D. Chong, and F. Nori, ``Edge Modes, Degeneracies, and Topological Numbers in Non-Hermitian Systems,"
Phys. Rev. Lett. \textbf{118}, 040401 (2017).

\bibitem{Kohmoto} K. Esaki, M. Sato, K. Hasebe, and M. Kohmoto, ``Edge states and topological phases in non-Hermitian systems," Phys. Rev. B \textbf{84}, 205128 (2011).

\bibitem{defectState} B. Qi, L. Zhang, and L. Ge, ``Defect States Emerging from a Non-Hermitian Flatband of Photonic Zero Modes," Phys. Rev. Lett. \textbf{120}, 093901 (2018).

\bibitem{NHFlatband_PRJ} L. Ge, ``Non-Hermitian lattices with a flat band and polynomial power increase [Invited]," Photon. Res. \textbf{6}, A10--A17 (2018).

\bibitem{Bender1} C.~M.~Bender and S.~Boettcher, ``Real spectra in non-Hermitian hamiltonians having $\cal PT$ symmetry," Phys. Rev. Lett. {\bf 80}, 5243 (1998).

\bibitem{Kawabata} K. Kawabata, S. Higashikawa, Z. Gong, Y. Ashida, and M. Ueda, ``Topological unification of time-reversal and particle-hole symmetries in non-Hermitian physics," Nat. Commun. \textbf{10}, 297 (2019).

\bibitem{Hatano} N. Hatano and D. R. Nelson, ``Localization Transitions in Non-Hermitian Quantum Mechanics," Phys. Rev. Lett. \textbf{77}, 570 (1996).

\bibitem{EP1} J. Okolowicz, M. Ploszajczak, and I. Rotter,
``Dynamics of quantum systems embedded in a continuum,"
Phys. Rep. {\bf 374}, 271 (2003).

\bibitem{EP2} W. D. Heiss, ``Exceptional points of non-Hermitian operators,"
J. Phys. A: Math. Gen. {\bf 37}, 2455 (2004).

\bibitem{EP3} A. U. Hassan, B. Zhen, M. Soljacic, M. Khajavikhan, and D. N. Christodoulides, ``Dynamically Encircling Exceptional Points: Exact Evolution and Polarization State Conversion," Phys. Rev. Lett. \textbf{118}, 093002 (2017).

\bibitem{EP4} C. Dembowski, H.-D. Gr\"af, H. Harney, A. Heine, W. Heiss, H. Rehfeld, and A. Richter,
``Experimental observation of the topological structure of exceptional points,"
Phys. Rev. Lett. {\bf 86}, 787 (2001).

\bibitem{EP5} S.-B. Lee, J. Yang, S. Moon, S.-Y. Lee, J.-B. Shim, S. Kim, J.-H. Lee, and K. An,
``Observation of an exceptional point in a chaotic optical microcavity,"
Phys. Rev. Lett. {\bf 103}, 134101 (2009).

\bibitem{EP6} M.~Liertzer, L.~Ge, A.~Cerjan, A.~D.~Stone, H.~E.~T\"{u}reci, and S.~Rotter,
``Pump-induced exceptional points in lasers,"
Phys. Rev. Lett. {\bf 108}, 173901 (2012).

\bibitem{EP_CMT} R.~El-Ganainy, M.~Khajavikhan, and L.~Ge, ``Exceptional points and lasing self-termination in photonic molecules," Phys.~Rev.~A \textbf{90}, 013802 (2014).

\bibitem{EP_ring} B. Zhen et al., ``Spawning rings of exceptional points out of Dirac cones," Nature \textbf{525}, 354 (2015).


%\bibitem{Song} A. Y. Song, Y. Shi, Q. Lin, and S. Fan, ``Direction-Dependent Parity-Time Phase Transition and Non-Reciprocal Directional Amplification with Dynamic Gain-Loss Modulation,'' Phys. Rev. A \textbf{99}, 013824 (2019).

%\bibitem{pseudoHermiticity1} A. Mostafazadeh, ``Pseudo-Hermiticity versus PT symmetry: The necessary condition for the reality of the spectrum of a non-Hermitian Hamiltonian," J. Math. Phys. \textbf{43}, 205 (2002).

%\bibitem{pseudoHermiticity2} A. Mostafazadeh, ``Pseudo-Hermiticity versus PT-symmetry. II. A complete characterization of non-Hermitian Hamiltonians with a real spectrum," J. Math. Phys. \textbf{43}, 2814 (2002).


\bibitem{Weinberg} S. Weinberg, \textit{The Quantum Theory of Fields, Vol. 1: Foundations} (Cambridge University, New York, 2005)

%\bibitem{Collins} M. J. Collins, F. Zhang, R. Bojko, L. Chrostowski, and M. C. Rechtsman, ``Integrated optical Dirac physics via inversion symmetry breaking," Phys. Rev. A \textbf{94}, 063827 (2016).

%\bibitem{Noh} J. Noh, S. Huang, K. P. Chen, and M. C. Rechtsman, ``Observation of Photonic Topological Valley Hall Edge States," Phys. Rev. Lett. \textbf{120}, 063902 (2018).


\bibitem{PT_PRX14} L. Ge and A. D. Stone, ``Parity-Time Symmetry Breaking beyond One Dimension: The Role of Degeneracy," Phys. Rev. X \textbf{4}, 031011 (2014).
\bibitem{conservation2D} L. Ge, K. G. Makris, D. N. Christodoulides, and L. Feng, ``Scattering in PT- and RT-symmetric multimode waveguides: Generalized conservation laws and spontaneous symmetry breaking beyond one dimension," Phys. Rev. A. \textbf{92}, 062135 (2015).
\bibitem{scattering_PRJ} L. Ge, ``Constructing the scattering matrix for optical microcavities as a nonlocal boundary value problem," Photon. Res. \textbf{5}, B20 (2017).

\bibitem{SSH} W. P. Su, J. R. Schrieffer, and A. J. Heeger, ``Solitons in polyacetylene," Phys. Rev. Lett. \textbf{42}, 1698 (1979).

\bibitem{Graefe} G. Demange and E.-M. Graefe, ``Signatures of three coalescing eigenfunctions," J. Phys. A: Math. Theor. \textbf{45}, 025303 (2012).
\bibitem{Flatband_PT} L. Ge, ``Parity-time symmetry in a flat-band system," Phys. Rev. A \textbf{92}, 052103 (2015).
\bibitem{Loncar} Z. Lin, A. Pick, M. Lon\u{c}ar, and A. W. Rodriguez, ``Enhanced Spontaneous Emission at Third-Order Dirac Exceptional Points in Inverse-Designed Photonic Crystals," Phys. Rev. Lett. \textbf{117}, 107402 (2016).
\bibitem{sensingEP3} H. Hodaei et al., ``Enhanced sensitivity at higher-order exceptional points," Nature \textbf{548}, 187 (2017).

%\bibitem{Nonl_PT} L. Ge and R. El-Ganainy, ``Nonlinear modal interactions in parity-time (PT) symmetric lasers," Sci. Rep. \textbf{6}, 24889 (2016).

%\bibitem{EP8} L. Ge, Y. D. Chong, S. Rotter, H. E. T\"{u}reci, and A. D. Stone, ``Unconventional modes in lasers with spatially varying gain and loss," Phys. Rev. A {\bf 84}, 023820 (2011).

%\bibitem{antiPT} L. Ge and H. E. T\"ureci, ``Antisymmetric PT-photonic structures with balanced positive- and negative-index materials," Phys. Rev. A {\bf 88}, 053810 (2013).

%\bibitem{antiPT_exp} P. Peng, W. Cao, C. Shen, W. Qu, J. Wen, L. Jiang, and Y. Xiao, ``Anti-parity-time symmetric optics via flying atoms," Nat. Phys. \textbf{3842} (2016).

\bibitem{Zhou} X. Zhou et al., ``Optical lattices with higher-order exceptional points by non-Hermitian coupling," Appl. Phys. Lett. \textbf{113}, 101108 (2018).

\bibitem{Rice-Mele} M. J. Rice and E. J. Mele, ``Elementary Excitations of a Linearly Conjugated Diatomic Polymer," Phys. Rev. Lett. \textbf{49}, 1455 (1982).
\bibitem{squareRoot} J. Arkinstall, M. H. Teimourpour, L. Feng, R. El-Ganainy, and H. Schomerus, ``Topological tight-binding models from nontrivial square roots," Phys. Rev. B \textbf{95}, 165109 (2017).

\bibitem{Redding} B. Redding, L. Ge, Q. Song, J. Wiersig, G. S. Solomon, and H. Cao, ``Local Chirality of Optical Resonances in Ultrasmall Resonators," Phys. Rev. Lett. \textbf{108}, 253902 (2012).
\bibitem{Sarma} R. Sarma, L. Ge, J. Wiersig, and H. Cao, ``Rotating Optical Microcavities with Broken Chiral Symmetry," Phys. Rev. Lett. \textbf{114}, 053903 (2015).
\bibitem{Cao} Q.-T. Cao et al., ``Experimental Demonstration of Spontaneous Chirality in a Nonlinear Microresonator," Phys. Rev. Lett. \textbf{118}, 033901 (2017).
\bibitem{Liu} S. Liu, J. Wiersig, W. Sun, Y. Fan, L. Ge, J. Yang, S. Xiao, Q. Song, and H. Cao, ``Transporting the Optical Chirality through the Dynamical Barriers in Optical Microcavities," Laser Photon. Rev. \textbf{12}, 1800027 (2018).

\end{thebibliography}
\end{document}